\documentclass[useAMS,usenatbib,usegraphicx]{mn2e}
\usepackage{amsmath,amssymb}
\usepackage{color}
\usepackage[english]{babel}
\def\ub {{\bf u}}

\def\Msyr {{\mathrm{M}_\odot\mathrm{yr}^{-1}}}
\def\df {{\mathrm{d}}}
\def\kms {{\mathrm{kms^{-1}}}}
\def\pc {{\mathrm{pc}}}
\def\kmss {{\mathrm{km^2s^{-2}}}}
\defcitealias{BS14}{BS14}
\defcitealias{FedKless12}{FK12}
\defcitealias{Krumholz2005}{KM05}
\defcitealias{PadNord09}{PN11}
\defcitealias{Padoan2012}{PHN12}
\defcitealias{SchmFed10}{SF11}
\defcitealias{BS12}{BS12}
\defcitealias{Schmidt2014}{SA14}
\defcitealias{HC08}{HC08-13}

\title[How the star formation law affects galactic disk structure]{The small 
and the beautiful: How the star formation law affects galactic disk structure}
\author[H.~Braun and W.~Schmidt]{H.~Braun$^{1}$\thanks{E-mail:
hbraun@astro.physik.uni-goettingen.de} and W.~Schmidt$^{1,2}$\thanks{E-mail:
wolfram.schmidt@uni-hamburg.de}\\
$^{1}$Institut f\"ur Astrophysik, Universit\"at G\"ottingen, Friedrich-Hund 
Platz 1, D-37077 G\"ottingen, Germany\\
$^{2}$Hamburger Sternwarte, Universit\"at Hamburg, Gojenbergsweg 112, D-21029 
Hamburg, Germany}

\begin{document}
\date{Draft version, April 2015}
\pagerange{\pageref{firstpage}--\pageref{lastpage}} \pubyear{????}
\maketitle
\label{firstpage}
\begin{abstract}
We investigate the influence of different analytical parameterizations and fit 
functions for the local star formation
rate in AMR simulations of an isolated disk galaxy with the \texttt{Nyx} code. 
Such parameterizations express the
star formation efficiency as function of the local turbulent Mach number and 
virial parameter. By employing the method
of adaptively refined large eddy simulations, we are able to evaluate these 
physical parameters from the numerically
unresolved turbulent energy associated with the grid scale. 
We consider both single and multi free-fall variants of star formation laws 
proposed by Padoan \& Nordlund, Hennebelle \& Chabrier,
and Krumholz \& McKee, summarised and tested recently with numerical simulations by Federrath \& Klessen.
We find that the global star formation rate and the relation between the local 
star formation rate
and the gas column density is reproduced in agreement with observational 
constraints by all multi free-fall models of
star formation. Some models with obsolete calibration or a single free-fall 
time scale, however, result in an overly
clumpy disk that does not resemble the structure of observed 
spirals.
\end{abstract}
\section{Introduction}\label{sec:intro}
Numerical simulations of disk galaxies cannot fully resolve processes in the 
interstellar medium (ISM) such as star formation and feedback from supernova 
(SN) explosions. For this reason, effective models for these processes are used 
in large-scale
simulations, which are called sub-resolution or sub-grid scale (SGS) models. 
The most commonly applied model of star formation is based on the
empirical Kennicutt-Schmidt (KS) relation 
\citep[e.g.][]{Schmidt1959,Kennicutt1998,GaoSol2004,Daddi2010,Genzel2010,
Lada2010,Bigiel2011,Kennicut2012,Schruba2013,Roychowdhury2015}. Formulated as 
an SGS model, this relation translates to
\begin{equation}
	\dot{\rho}_{\rm s} = \epsilon_{\rm ff}\frac{\rho}{\tau_{\rm ff}},
\end{equation}
where $\rho$ can be either the total gas density 
\citep[e.g.][]{Renaud2013,Agertz2013,Lagos2015,Tasker2015} or some fraction of 
that density (for example, the molecular
hydrogen density as in \citet{Gnedin2009}, \citet{Dobbs2013}, or 
\citet{Agertz2014}), $\tau_{\rm ff}\sim (G\rho)^{-1/2}$ is the free-fall time 
scale associated with $\rho$, and $\epsilon_{\rm ff}$ the so-called star 
formation efficiency. The coefficient $\epsilon_{\rm ff}$ is usually assumed to
be a constant, i.~e.\ a freely tunable numerical parameter, in disk galaxy 
simulations. This constant is chosen such that
the global star formation rate matches observational values. However, numerical 
studies and observations of star-forming regions
in galaxies suggest significant local variations of the star formation 
efficiency \citep{Evans2009,Onodera2010,Murray2011,Federrath2013,Salim2015}, which may arise from 
different evolutionary stages of local star-forming environments \citep{Kruijssen2014}.

Also from the theoretical point of view, the star formation efficiency should 
change with the local conditions in the ISM because
stars form from gravitationally unstable density enhancements seeded by 
turbulence in molecular clouds \citep[e.g.][]{LowKless04,McKee2007,HenneFal12}. 
Although the dynamics of star-forming clouds appears to be extremely complex 
(compressible turbulence, self-gravity, magnetic fields, radiative cooling and 
heating), attempts have been made to capture the essence of the star formation 
process in relatively simple analytical relationships between the star 
formation rate or efficiency and some fundamental parameters. 
In particular, star formation efficiencies were parameterized in terms of the 
turbulent Mach number
(corresponding to the ratio of turbulent and thermal energies) and the viral 
parameter (the ratio of turbulent and gravitational
energies). An representative example is the relation proposed by 
\citet{PadNord09} 
(a complete overview will be given in Section~\ref{sec:methods}),
which is based on the idea that the typical length scale of dense clumps is 
given by the thickness of shock layers for
a particular Mach number of turbulence and that the clumps become unstable if 
they exceed their Jeans mass 
(this is where the virial parameter comes in). The distribution of density 
enhancements is determined by a log-normal density distribution, which again 
depends on the turbulent Mach number \citep{Federrath2015}. \citet{BS14} [hereafter 
\citetalias{BS14}] exploited
this relation in order to dynamically compute the efficiency parameter in 
simulations of an isolated disk galaxy, rather
than assuming some specific value. It turned out that this kind of SGS model is 
able to predict typical
star formation efficiencies in galaxies 
\citep[e.g.][]{GaoSol2004,Daddi2010,Genzel2010,Murray2011} 
without making use of the observed relations for the star formation rate as 
input.
However, an open question is whether the properties of simulated disk galaxies 
are sensitive to the
various assumptions that go into the star formation model. If that were the 
case, it should be easily possible to discriminate
existing models, which make entirely different basic assumptions. On the other 
hand, it is also possible that different models
produced comparable effects under conditions that favour star formation.

Another essential component of a SGS model for realistic galaxy simulations is 
stellar feedback. 
There is a great variety of different approaches and extensive studies of the 
influence of feedback models
on the properties of simulated galaxies 
\citep[e.g.][]{StinSeth06,Stinson2013,Wise2012,Agertz2013,Hopkins2014}. 
Here we focus on the influence of the star formation model on global properties 
of simulated isolated disk galaxies for one particular feedback model, namely 
the mixed thermal and turbulent feedback introduced in \citetalias{BS14}. It 
was demonstrated that
this model produces reasonable results and therefore can be considered as a 
reliable reference point for further studies. 

This article is structured as follows. After outlining the applied numerical 
methods and summarising the star formation
models implemented in our code (Section~\ref{sec:methods}), we present the 
results from our simulations in Section~\ref{sec:result}.
In particular, we consider the global star formation rate, the structure of the 
gaseous disk, the relations between the local star
formation rate and surface densities, and the role of turbulence for the 
different models. The results of our study are 
summarised and discussed in Section~\ref{sec:conclude}. 

\section{Numerical methods and models}\label{sec:methods}

The simulations presented in this study were carried out using the cosmological 
hydrodynamics code \texttt{Nyx} \citep{NYX}. \texttt{Nyx} solves the Euler 
equations on an adaptively refined grid using the piecewise parabolic method. 
Collisionless components like stellar particles are taken into account as 
N-body 
system via a particle-mesh approach. In our simulations, we included the SGS 
model introduced by \citet{SchmFed10} and \citet{Schmidt2014} to model 
turbulent 
motions below the resolution scale. Non-adiabatic sub-resolution processes are 
modelled using the MIST (Multi-phase Interstellar medium with Star formation 
and 
Turbulence) model, which was proposed in \citet[][hereafter \citetalias{BS12}]{BS12} and \citetalias{BS14}. 
The 
employed physics encompasses:
\begin{itemize}
 \item \textit{Self gravity} from gas and stars.
 \item \textit{Gas dynamics} using piecewise parabolic method for the resolved 
motions and the SGS model for unresolved turbulence.
 \item \textit{Radiative cooling} from gas, metals, and dust.
 \item \textit{Multi-phase ISM} consisting of a diffuse warm, a clumpy cold, 
and a hot phase for SN ejecta.
 \item \textit{Star formation rate} depending on the locally inferred molecular 
fraction and the thermal and turbulent state of the gas.
 \item \textit{Stellar feedback} in the form of a combination of thermal and 
turbulent feedback from SN and thermal feedback from Lyman continuum heating, 
both depending on the age of the stars.
\item \textit{Metal enrichment} due to SN.
\end{itemize}
As demonstrated in \citetalias{BS14}, simulations featuring MIST in combination 
with a SGS turbulence model are able to simultaneously reproduce several 
observed properties of star formation in a plausible way. Owing to the combined 
effect of turbulent plus thermal SN-feedback and a non-cooling, but decaying 
hot 
gas-phase of SN-ejecta the so-called over-cooling problem is avoided. The star 
formation relations of KS-type following from the simulation data are a genuine 
result of the modelling, as they were not a priori imposed in the model. 
Moreover, MIST allows us to probe the behaviour of different theoretical 
approaches to the calculation of the star formation rate in the cold molecular 
gas by incorporating the corresponding parameterizations in our simulations. 

\subsection{Multi-phase and star formation 
models}\label{sec:sfr}

MIST separates the gas content of a numerical resolution element into two 
fractions -- or phases: a clumpy cold phase and a diffuse warm phase, within 
which the cold clumps are embedded.  
The fractional densities of the cold and warm phases are 
determined by partial differential equations with source terms related to 
various exchange processes between the phases and star formation, which 
converts cold 
gas into stellar mass (see Figures~1 and~2 of \citetalias{BS12} for an 
overview). 
While the warm phase is subject to radiative cooling and feeds the clumpy cold 
phase through this channel, the cold phase itself is kept at a constant 
temperature of $T_{\rm c}=50\ \mathrm{K}$. The heating of cold gas 
is modelled as mass transfer into the warm phase, which has a variable 
temperature following 
form the hydrodynamical energy equation. Under adiabatic conditions, the total 
energy of the two phases is conserved.
Moreover, shielded molecular gas may reside inside the cold clumps. The mass 
fraction $f_{\rm H_2}$ of 
molecular gas is estimated using a Str\"omgren-like approach.

Only this molecular fraction $f_{\rm H_2}\rho_{\rm c}$ of the cold gas with fractional 
density $\rho_{\rm c}$ is allowed to form stars at a rate $\dot{\rho}_{\rm s}$ 
given by
\begin{equation}
 \label{eq:sf}
 \dot{\rho}_{\rm s}=\dfrac{\epsilon_{\rm core}\epsilon_{\rm mod}f_{\rm 
H_2}\rho_{\rm c}}{\tau_{\rm c,ff}}.
\end{equation}
$\tau_{\rm c,ff}$ is the free-fall time scale within the cold clumps
\begin{equation}
  \label{eq:tau_ff}
 \tau_{\rm c,ff}=\displaystyle{\sqrt{\frac{3\mathrm{\pi}}{32G\rho_{\rm c,pa}}}},
\end{equation}
with the gravitational constant $G$ and the average density inside the cold 
phase $\rho_{\rm c,pa}$, which differs from the fractional density $\rho_{\rm 
c}$. To calculate the average densities, an equilibrium between the effective 
(thermal plus turbulent) pressures of the cold and warm phases is assumed at 
the 
typical clump scale $\ell_{\rm c}$ \citepalias[for a thorough description 
see][]{BS12,BS14}. 
The coefficient $\epsilon_{\rm mod}$ in equation~(\ref{eq:sf}) specifies the
efficiency of forming gravitationally bound cores with respect to the free-fall 
time scale $\tau_{\rm c,ff}$. The core efficiency parameter $\epsilon_{\rm core}=0.4$ 
takes into account that only a fraction of the mass in the bound cores eventually 
ends up in stars. The remaining fraction 
$(1-\epsilon_{\rm core})$ of the prestellar material is assumed to be re-injected into the ISM 
due to protostellar jets, winds, and outflows.

No explicit density threshold is imposed on the 
total gas density $\rho$ or the average density of the cold phase, $\rho_{\rm 
c,pa}$. The only requirement for star formation is the existence of 
shielded molecular gas inside the cold clumps, which depends on $\rho_{\rm 
c,pa}$, $\ell_{\rm c}$, and the metallicity $Z$.

A unique feature of MIST is that $\epsilon_{\rm mod}$ is not assumed to 
be a constant coefficient, but is treated as a variable of the model, which is 
computed from the current local state of the gas in the cold clumps. 
This is achieved by modelling the substructure of the ISM and turbulence
below the resolution scale on the basis of the numerically unresolved
turbulent energy $K_{\rm SGS}$, which determines key parameters in the
expressions for $\epsilon_{\rm mod}$ following from analytical theories
of star formation.

In \citetalias{BS12} and \citetalias{BS14}, the model of \citet[][hereafter 
\citetalias{PadNord09}]{PadNord09} in the parameterization of \citet[][in the 
following \citetalias{FedKless12}]{FedKless12} were used. In this study, we 
explore the effects of different choices for $\epsilon_{\rm mod}$
based on the analytical models proposed by \citetalias{PadNord09}, 
\citet[][hereafter \citetalias{Krumholz2005}]{Krumholz2005}, and 
\citet[][hereafter \citetalias{HC08}]{HC08,HC09,HenChab11,HC13}. While these 
models result from theoretical reasoning, a simple star formation law was put 
forward by 
\citet[][hereafter \citetalias{Padoan2012}]{Padoan2012} by fitting data from 
simulations of driven isothermal turbulence in self-gravitating gas.

In the context of the analytical models, $\epsilon_{\rm mod}$ can be understood 
as an $x$-weighted integral of the probability density function 
$\mathrm{pdf}(x)$ over the logarithmic overdensity ratio 
$x=\mathrm{log}(\tilde{\rho}/\rho_{\rm c,pa})$ above a certain critical 
threshold 
$x_{\rm cr}=\mathrm{log}(\tilde{\rho}_{\rm cr}/\rho_{\rm c,pa})$, which 
separates gravitationally bound from unbound structures in the 
turbulent 
gas. To account for a collapse time scale that varies with the
density fluctuation $x$, multi free-fall (mff) models incorporate the weighting 
factor 
$f_{\rm ff}=\tau_{\rm c,ff}/\tau_{\rm ff}(\tilde{\rho})=\mathrm{exp}^{1/2}(x)$ into 
the integral.  
In general, $\epsilon_{\rm mod}$ can be written as
\begin{equation}
 \epsilon_{\rm mod}=f_{\rm cal}\displaystyle{\int^\infty_{x_{\rm cr}}}f_{\rm 
ff}(x)\;\mathrm{exp}(x)\;\mathrm{pdf}(x)\;\df x.
\end{equation}
The arbitrary prefactor $f_{\rm cal}$ ($1/\phi_{t}$ in \citetalias{FedKless12})
has to be calibrated by fitting the star formation efficiency to numerical data
from simulations of self-gravitating turbulence.

Although $\mathrm{pdf}(x)$ tends to develop a power-law tail in strongly 
self-gravitating gas \citep[e.g.][]{Girichidis2014,Kainulainen2009}, the analytical models generally assume a log-normal 
shaped PDF (the reasons of which are discussed in \citet{FedKless13}), which is characteristic for supersonic isothermal turbulence: 
\begin{equation} 
\mathrm{pdf}(x)=\dfrac{1}{\sqrt{2\pi\sigma_{x}^2}}\;\mathrm{exp}\!\left(-\dfrac{
(x-x_0)^2}{2\sigma_x^2}\right),
\end{equation}
where 
\begin{equation}
 \label{eq:sigma_x}
 \sigma_x^2=\mathrm{log}\left(1+b^2\mathcal{M}^2\right),
\end{equation}
determines the width of the distribution and $x_0=-\sigma_x^2/2$ is the 
logarithmic mean.
Apart from the compressive factor\footnote{
This definition assumes that turbulence 
injection driven due to SNe is dominated by compressive/dilatational forces. 
Otherwise a 
mixture of compressive and solenoidal driving forces resulting from shearing 
motions in the 
galactic disk, local gravitational collapse, or motions triggered by thermal 
instabilities of the multi-phase ISM is assumed.
}
\begin{equation}
 b=\begin{cases}
   1&\mbox{if local SNe feedback is active}\\
   0.5&\mbox{in quiescent regions}
   \end{cases},
\end{equation}
which was introduced by \citet{Federrath2010b}, $\sigma_x$ depends on the 
root-mean-square Mach number $\mathcal{M}$ of turbulent motions in the cold 
clumps. We estimate $\mathcal{M}$ in our simulations by extrapolating the 
turbulent SGS energy from the grid scale to the clump scale
$\ell_{\rm c}$ by assuming a power law for the scale dependence of turbulent 
velocity fluctuations:
\begin{equation}\label{eq:M_c}
 \mathcal{M}=\displaystyle{\sqrt{\dfrac{2K_{\rm SGS}\left(\ell_{\rm 
c}/\Delta\right)^{2\eta_{\rm w}}}{\gamma(\gamma-1)e_{\rm c}}}}
\end{equation}
with the specific thermal energy of the cold phase $e_{\rm c}=e(T_{\rm c}=50\ 
\mathrm{K})$, the adiabatic index of the equation of state of the gas 
$\gamma=5/3$, the specific kinetic energy $K_{\rm SGS}$ of motions below 
resolution scale $\Delta$, and the turbulent velocity scaling exponent 
$\eta_{\rm w}=1/3$.

Basically, the differences between the analytical models amount to different 
definitions of 
$x_{\rm cr}$ and $f_{\rm ff}$. The critical overdensity $x_{\rm cr}$ is in all 
cases a function of $\mathcal{M}$ and the virial parameter $\alpha_{\rm vir}$ 
of the cold clumps, where
\begin{equation}\label{eq:alpha_vir}
 \alpha_{\rm vir}=\dfrac{10K_{\rm SGS}\left(\ell_{\rm 
c}/\Delta\right)^{2\eta}}{\mathrm{\pi} G\ell_{\rm c}^2\rho_{\rm c,pa}}.
\end{equation}

In addition to the models in their originally published form, we also consider 
the single and multi free-fall formulations with the best-fit calibrations as 
provided by \citetalias{FedKless12} 
in their tables~1 and~3. A comprehensive list of all the models used in this 
study is given in Table~\ref{tab:SFR_mods}.
The main ideas behind these models are briefly summarised in the following.
For a more detailed account see \citetalias{FedKless12} and the original
papers.\footnote{
In the following we refer to different subsets out of our simulation suite as 
families. PN, PN-FK, and PN-FK-mff belong to the PN-family, HC, HC-FK, and 
HC-FK-mff to the HC-family, and KM, KM-FK, and KM-FK-mff accordingly to the 
KM-family.}

\begin{table*}\begin{minipage}{177mm}
\caption{Overview of star formation models implemented into 
MIST. 
For a brief description of the models and definitions of the various 
coefficients
and parameters, see Sections~\ref{sc:PN_family}--\ref{sc:KM_family}.}
 \centering
\label{tab:SFR_mods}
 \begin{tabular}{l|l|l|l|l|l}
  \multicolumn{6}{c}{ }\\
  \hline
  ID&critical overdensity ratio&ff-time factor&SF 
efficiency&coefficients&references\\
  &$\mathrm{exp}(x_{\rm cr})$&$f_{\rm ff}$&$\epsilon_{\rm mod}$&&\\
  \hline
  \hline
  \multicolumn{6}{c}{models and coefficients as in original publication (see 
listed reference -- if they did not suggest a value, we chose $f_{\rm cal}=1$)} 
\\
  \hline
  PN&$(0.0067)\theta^{-2}\alpha_{\rm 
vir}\mathcal{M}^{2}$&$\mathrm{exp}^{1/2}(x_{\rm cr})$&$f_{\rm cal}\times 
a_2$&$\theta=0.5$, $f_{\rm cal}=1.0$&\citetalias{PadNord09}\\
  HC&$(\pi^2/5)y_{\rm cut}^{-1}\alpha_{\rm vir}\left(y_{\rm 
cut}^{-1}\mathcal{M}^{-2}+1/3\right)$&$\mathrm{exp}^{1/2}(x)$&$f_{\rm 
cal}\times a_3$&$y_{\rm cut}=0.1$, $f_{\rm cal}=1.0$&\citetalias{HC08}\\
  KM&$(\pi^2/5)\phi_x^2\alpha_{\rm vir}\mathcal{M}^2$&1&$f_{\rm cal}\times 
a_1$&$\phi_x=1.12$, $f_{\rm cal}=1.92$&\citetalias{Krumholz2005}\\
  PHN&n.a.&n.a.&$\mathrm{exp}\!\left[\varphi\pi^2\alpha_{\rm 
vir}\right]$&$\varphi=-0.003$&\citetalias{Padoan2012}\\
  \hline
  \multicolumn{6}{c}{calibrated by \citetalias{FedKless12} (see their table~1 
for analytic forms and their table~3 (HD-fit) for the coefficients)} \\
  \hline
  PN-FK&$(0.0067)\theta^{-2}\alpha_{\rm 
vir}\mathcal{M}^{2}$&$\mathrm{exp}^{1/2}(x_{\rm cr})$&$f_{\rm cal}\times 
a_2$&$\theta=1.0$, $f_{\rm cal}=1.5$&\citetalias{PadNord09,FedKless12}\\
  HC-FK&$(\pi^2/5)y_{\rm cut}^{-1}\alpha_{\rm vir}\left(y_{\rm 
cut}^{-1}\mathcal{M}^{-2}+1/3\right)$&$\mathrm{exp}^{1/2}(x)$&$f_{\rm 
cal}\times a_3$&$y_{\rm cut}=1.3$, $f_{\rm 
cal}=0.24$&\citetalias{HC08,FedKless12}\\
  KM-FK&$(\pi^2/5)\phi_x^2\alpha_{\rm vir}\mathcal{M}^2$&1&$f_{\rm cal}\times 
a_1$&$\phi_x=0.12$, $f_{\rm cal}=3.0$&\citetalias{Krumholz2005,FedKless12}\\
  \hline
  \multicolumn{6}{c}{multi free-fall (mff), calibrated by 
\citetalias{FedKless12} (see their table~1 for analytic forms and table~3 
(HD-fit) for the coefficients)} \\
  \hline
  PN-FK-mff&$(0.0067)\theta^{-2}\alpha_{\rm 
vir}\mathcal{M}^{2}$&$\mathrm{exp}^{1/2}(x)$&$f_{\rm cal}\times 
a_3$&$\theta=1.0$, $f_{\rm cal}=0.49$&\citetalias{PadNord09,FedKless12}\\
  HC-FK-mff&$(\pi^2/5)y_{\rm cut}^{-2}\alpha_{\rm 
vir}\mathcal{M}^{-2}$&$\mathrm{exp}^{1/2}(x)$&$f_{\rm cal}\times a_3$&$y_{\rm 
cut}=1.1$, $f_{\rm cal}=0.21$&\citetalias{HC08,FedKless12}\\
  KM-FK-mff&$(\pi^2/5)\phi_x^2\alpha_{\rm 
vir}\mathcal{M}^2$&$\mathrm{exp}^{1/2}(x)$&$f_{\rm cal}\times 
a_3$&$\phi_x=0.19$, $f_{\rm cal}=0.49$&\citetalias{Krumholz2005,FedKless12}\\
  \hline
  \hline\\
  \multicolumn{2}{l}{\textbf{Definitions: 
}}&\multicolumn{3}{l}{$a_1=0.5\left(1+\mathrm{erf}\!\left[(\sigma_x^2-2x_{\rm 
cr})/(8\sigma_x^2)^{1/2}\right]\right)$;}\\
  
\multicolumn{2}{l}{}&\multicolumn{3}{l}{$a_2=0.5\left(1+\mathrm{erf}\!\left[
(\sigma_x^2-2x_{\rm 
cr})/(8\sigma_x^2)^{1/2}\right]\right)\mathrm{exp}\left(x_{\rm cr}/2\right)$;}\\
  \multicolumn{2}{l}{ 
}&\multicolumn{4}{l}{$a_3=0.5\left(1+\mathrm{erf}\!\left[(\sigma_x^2-x_{\rm 
cr})/(2\sigma_x^2)^{1/2}\right]\right)\mathrm{exp}\!\left(3\sigma_{x}
^2/8\right)$}\\
  \end{tabular}
\end{minipage}\end{table*}

\subsubsection{PN family}
\label{sc:PN_family}
\citetalias{PadNord09} assume that the size of a critical Bonnor-Ebert-sphere 
is equal to the typical thickness of a shocked layer, which is computed from the 
combination of the isothermal shock jump conditions and a turbulent velocity 
scaling relation $u^\prime(\ell)\propto\ell^{\eta_{\rm c}}$ with $\eta_{\rm 
c}=0.5$. This 
implies
\begin{equation}
  \label{eq:x_cr_PN}
 \mathrm{exp}(x_{\rm cr})=0.0067\ \theta^{-2}\alpha_{\rm vir}\mathcal{M}^2.
\end{equation}
In the MIST framework, $\theta\ell_{\rm c}$ is the injection scale of supersonic
turbulence in a cloud.
In the case of our fiducial model PN-FK we simply set $\theta=1.0$, while 
$\theta=0.35$ in \citetalias{PadNord09} and $\theta=0.65$ in table~3 of 
\citetalias{FedKless12}. We chose the intermediate value $\theta=0.5$ for our 
PN run to explore the influence of this coefficient. Both the original 
\citetalias{PadNord09}
model and PN-FK assume a constant free-fall time factor 
$f_{\rm ff}=\mathrm{exp}^{1/2}(x_{\rm cr})$, i.e.\ the effective
free-fall time scale is given by the critical overdensity $\tilde{\rho}_{\rm cr}$ rather than
the mean density $\rho_{\rm c,pa}$ of the cold phase.
\subsubsection{HC family}
\label{sc:HC_family}
The model suggested by \citetalias{HC08} is based on the assumption that
bound substructures of a turbulent gas cloud collapse on time scales determined 
by the mean densities of the substructures rather than the whole cloud. 
The star formation efficiency is then obtained by integrating 
the mass spectrum of substructures weighted by their free-fall time 
factors. This integral is truncated at a certain mass fraction $y_{\rm cut}$ 
to avoid overly large structures. 
While \citetalias{HC08} suggested $y_{\rm cut}\approx0.1$,
\citetalias{FedKless12} concluded from their fits to simulation 
data that $y_{\rm cut}$ slightly above unity is favoured.\\
Assuming that additional non-thermal support against gravity must increase the 
stability range 
(a notion that was questioned by \citealt{SchmColl13}), the critical 
overdensity is defined by requiring 
that the turbulent Jeans length $\lambda_{\rm J,t}(\rho_{\rm cr})$ 
\begin{equation}
\lambda_{\rm J,t}(\rho_{\rm 
cr})=\displaystyle{\sqrt{\frac{\mathrm{\pi}\gamma(\gamma-1)e_{\rm 
c}+2\mathrm{\pi}\lambda_{\rm J,t}(\rho_{\rm cr})K_{\rm 
SGS}\left(\frac{\ell_{\rm 
c}}{\Delta}\right)^{2\eta_{\rm w}}}{G\rho_{\rm cr}}}}
\end{equation}
equals the size $y_{\rm cut}\ell_{\rm c}$ of the largest collapsing 
substructures. 
This condition leads to a quadratic equation in $\ell_{\rm c}$.
Substituting the definitions of $\mathcal{M}$ and $\alpha_{\rm vir}$ yields 
\begin{equation}
  \label{eq:x_cr_HC}  
 \mathrm{exp}(x_{\rm cr})=\dfrac{\mathrm{\pi}^2\alpha_{\rm vir}}{5y_{\rm 
cut}^2\mathcal{M}^2}+\dfrac{\mathrm{\pi}^2\alpha_{\rm vir}}{15y_{\rm cut}}.
\end{equation}
This expression is used for the models HC and HC-FK. Following 
\citetalias{FedKless12}, the second term is dropped for the
HC-FK-mff model. For the HC family, 'mff' is actually a misnomer because
all models in this family feature the multi free-fall time factor
$f_{\rm ff}=\mathrm{exp}^{-1/2}(x)$. 
\subsubsection{KM family}\label{sc:KM_family}
\citetalias{Krumholz2005} argue that the critical overdensity follows from the 
condition that the sonic length is close to the Jeans length. From this 
condition it follows that 
\begin{equation}
 \mathrm{exp}(x_{\rm cr})=(\mathrm{\pi}^2/5)\phi_x^2\alpha_{\rm vir}\mathcal{M}^2.
\end{equation}
Apart from the prefactor, the resulting expression for $\mathrm{exp}(x_{\rm 
cr})$ is the same as for the models in the PN family. 
The fudge factor $\phi_x$ should be of the order unity. 
\citetalias{Krumholz2005} obtain the best fit for $\phi_x=1.12$. However, 
\citetalias{FedKless12} determined smaller values 
($\phi_x\sim0.1\ldots0.2$).\\
The main difference between the original PN and KM models is that there is no
free-fall time factor at all in the KM models ($f_{\rm ff}=1$ 
in the case of KM and KM-FK). This means that \citetalias{Krumholz2005} assume that the free-fall
time scale is just given by the mean density of the cold gas (see equation~\ref{eq:tau_ff} for $\tau_{\rm c,ff}$), 
while PN and PN-FK set the effective free-fall time scale in the star formation law
equal to $\tau_{\rm ff}(\tilde{\rho}_{\rm cr})$ (in this case, $\tau_{\rm c,ff}$
cancels out from equation~(\ref{eq:sf}) for the star formation rate). 
The mff variants of the KM and PN models, on the other hand, are formally the same.
\subsection{Simulations}\label{sec:runs}
For each of the ten star formation models in Table~\ref{tab:SFR_mods} a 
simulation run of an idealised isolated disk galaxy was performed with 
\texttt{Nyx} \citep{NYX} using the methodology from \citepalias{BS14}.
As initial conditions we used data obtained from the 'ref' of \citetalias{BS14} run with their fiducial 
model (in this paper referred to as PN-FK) at a simulated age of roughly 1~Gyr. Starting 
the simulations for our comparison from a pre-evolved galaxy has several 
benefits. Employing an aged galaxy as initial conditions minimises transient 
effects of the rapid initial growth of the stellar mass and the metallicity 
within the galactic disk. Since we begin with a statistically stationary star 
forming configuration of the galactic disk, in which star formation and stellar 
feedback effects are nearly balanced, it is easy to detect the changes of the 
disk configuration after switching to a different star formation model.

The 'ref' run of \citetalias{BS14} simulated the evolution of an idealised 
isolated disk galaxy residing in a box of $(0.5\ \mathrm{Mpc})^3$ size 
with the PN-FK model. With a root grid of $256^3$ and 6 levels 
of adaptive mesh refinement an effective resolution of 30~pc was achieved. The 
galaxy in 'ref' was initialised as an adiabatically stable, rotating gas disk of $10^{10}\ 
\mathrm{M}_\odot$ mass using the potential-method described by 
\citet{WangKless10}. The static potential of an NFW-shaped dark matter halo of 
$~10^{12}\ \mathrm{M}_\odot$  is added to the self-gravity of the gaseous disk. 
There was no initial stellar component. Radiative cooling causes the initially
thick and hot disk to rapidly collapse into a fragmenting thin disk.

After $1\ \mathrm{Gyr}$ of evolution, about $30$ per cent of the initial 
total gas mass has been converted into stars. While the majority of the long-lived 
stars are rather smoothly distributed over the inner disk, many young stars 
reside in stellar clusters with a typical lifetime of a few 100~Myr. The gas 
content of the inner, star-forming part of the disk is below $50$ per cent. At this 
stage, the simulated galaxy roughly resembles a disk galaxy in a quiescent star 
forming state with a global star formation rate of $\sim1\ 
\mathrm{M_\odot}\mathrm{yr}^{-1}$ governed by the self-regulation 
between star formation and stellar feedback.\footnote{
Our configuration is not comparable to a star-burst galaxy at high redshift such 
as the 'HighZ' runs in \citet{Hopkins2011}. After $1\ \mathrm{Gyr}$, the 
conditions in the inner disk of our simulation roughly correspond to the 'Sbc' 
runs in that paper.
}
We use this configuration as initial condition for our comparison runs by following 
the disk evolution over additional 0.4~Gyr with different star formation models. 
The time period of 0.4~Gyr corresponds to one orbital revolution at 10~kpc 
distance from the galactic centre or roughly ten times the maximum age of a
stellar particle producing feedback due to SNe of type II.

\section{Results}\label{sec:result}
\subsection{Global star formation rate}
\label{sc:global_sf_rate}
\begin{figure}
\centering
  \includegraphics[width=0.95\hsize]{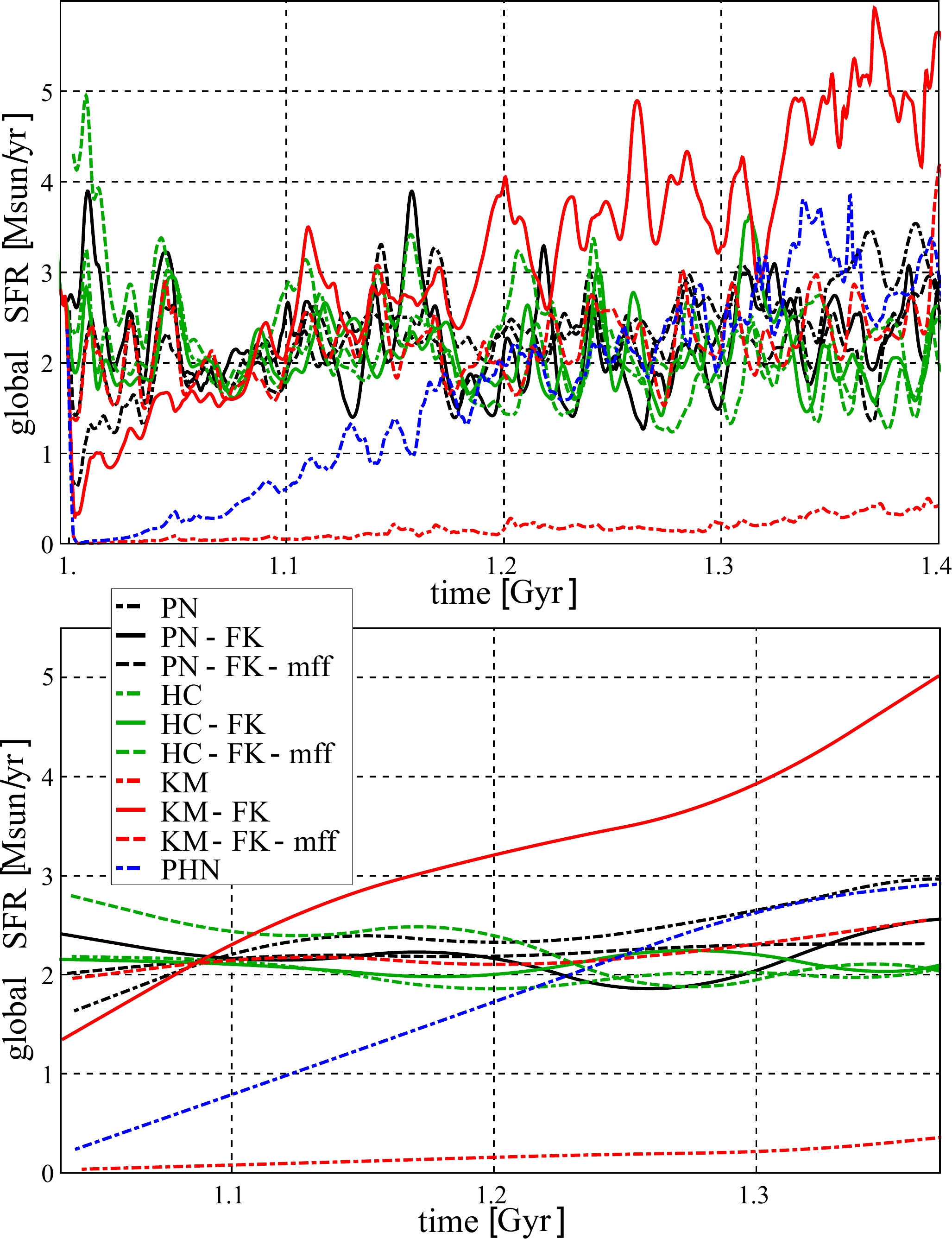}
\caption{ \textbf{Top panel: }Global star formation rate $\dot{M}_{\rm SF}$ 
versus simulation time for runs using the different star formation models 
listed in Table~\ref{tab:SFR_mods}. The evolution of $\dot{M}_{\rm SF}$ in runs 
with \citetalias{PadNord09}-family models is drawn in black, with 
\citetalias{HC08}-family models in green, with \citetalias{Krumholz2005}-family 
models in red, and with the \citetalias{Padoan2012} model in blue. A model in 
its original version is indicated by a dot-dashed line, its 
\citetalias{FedKless12}-calibrated version by a solid line, and its 
\citetalias{FedKless12}-calibrated multi free-fall version by a dashed line. 
\textbf{Bottom panel: }Same as above, but $\dot{M}_{\rm SF}$ is smoothed on a 
time scale of 80~Myr using a moving average to filter out short term 
variations.}
\label{fig:BS14a_SFRvsTime}
\end{figure}
The star formation rate integrated over the whole disk is plotted as function 
of time in Fig.~\ref{fig:BS14a_SFRvsTime} for all runs. After a short 
transitional phase, in which the disk configuration adjusts itself to the new 
conditions set by the star formation model, the global star formation rate 
$\dot{M}_{\rm SF}$ settles on average between $2\ \Msyr$ and $3\ \Msyr$ in the majority 
of the runs (PN-family, HC-family, and KM-FK-mff). In the bottom panel of 
Fig.~\ref{fig:BS14a_SFRvsTime}, short term variations are smoothed out by 
applying a moving average with width of $\sim$80~Myr. These variations, which 
can be as large as $1\ \Msyr$ in amplitude, are a consequence of the life-cycle 
of individual star-forming regions in the disks. 

As explained in \citetalias{BS14}, the associated time scale of around 
$10$ to $30\ \mathrm{Myr}$ arises from delayed SNe feedback.
This time scale is consistent with observational estimates of 
the life-time of star-forming gas clouds 
\citep[e.g.][]{Blitz2007,McKee2007,Miura2012} and is also reproduced by 
numerical simulations \citep[e.g.][]{Hopkins2012,Tasker2015,Dobbs2013}. 
The smoothed time evolution of $\dot{M}_{\rm SF}$ is similar for most runs, 
including the PN-FK run, which merely continues the 'ref' run of 
\citetalias{BS14}.

However, the runs KM, KM-FK, and PHN exhibit deviating behaviour. These three 
runs have in common that $\dot{M}_{\rm SF}$ drops well below $1\ \Msyr$ shortly 
after the start of the simulation and then gradually grows until the simulation 
is stopped. Only the PHN run with the simple star formation model seems to 
enter a self-regulated regime toward the end of the simulation, where 
$\dot{M}_{\rm SF}$ approaches a similar level as in the other simulations. 
While the KM-FK model leads to an overproduction of stars compared to the 
continuation  of the 'ref' run with the PN-FK model, 
the original KM model falls much below the expected star formation rate. The 
deviations observed for the KM approach are remedied with the multi free-fall 
modification of the star formation law. Otherwise, the global star formation 
rate appears to be quite robust with respect to the parameterization of star 
formation.
\subsection{Turbulence}
\begin{figure}
\centering
  \includegraphics[width=0.95\hsize]{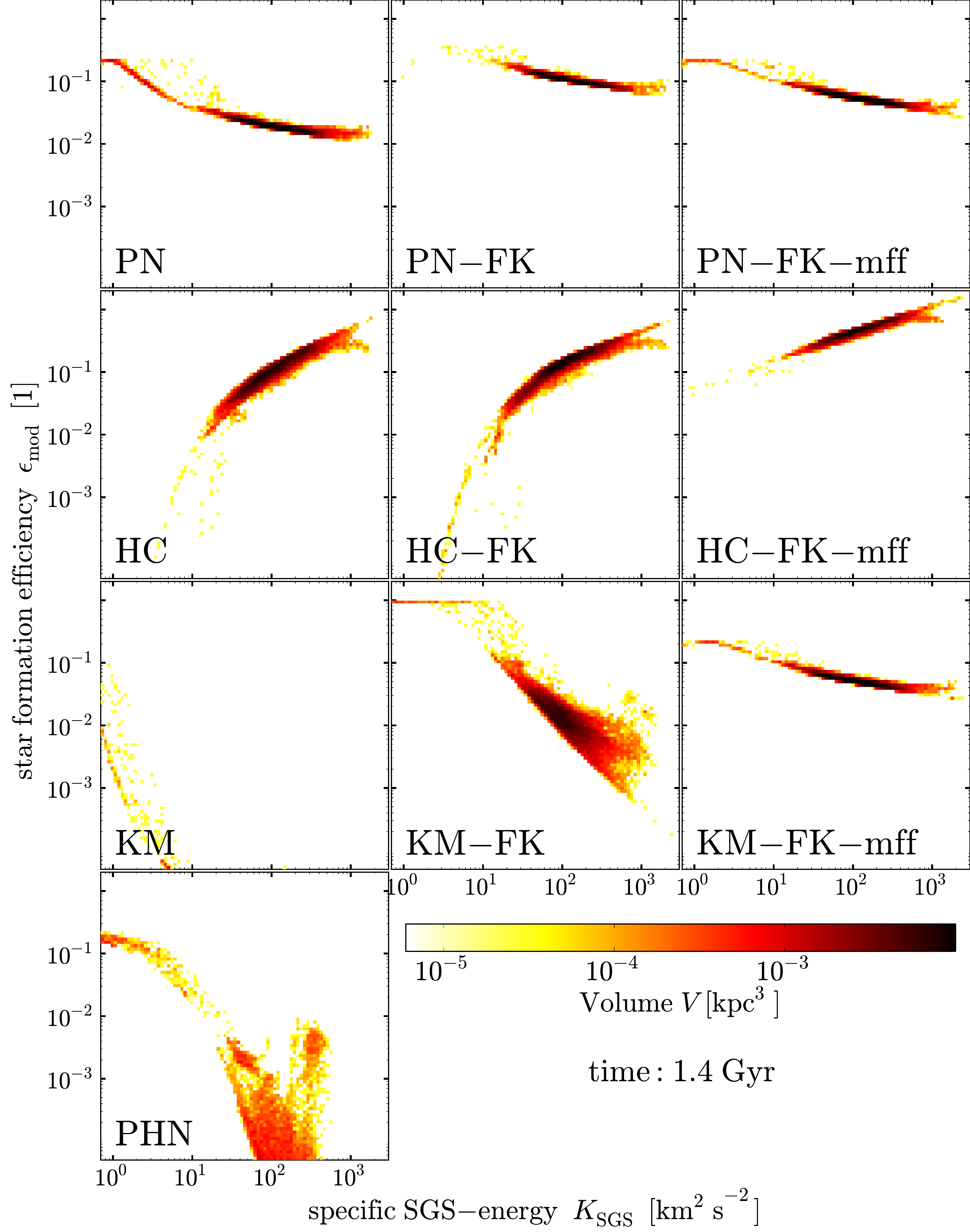}
\caption{Comparison of area-weighted 2D histograms of the star formation 
efficiency $\epsilon_{\rm mod}$ in the cold phase vs.\ the SGS turbulence 
energy $K_{\rm SGS}$ 
at an age of $\sim1.4$~Gyr. The area is logarithmically colour-coded.}
\label{fig:BS14a_synopsis_K_vs_EffC}
\end{figure}
Figure~\ref{fig:BS14a_synopsis_K_vs_EffC} reveals that star formation occurs for all included models preferentially around the sweet 
spot at specific SGS energies of $K_{\rm SGS}\sim 100\ 
\mathrm{km}^2\mathrm{s}^{-1}$.
This corresponds to a preferred velocity dispersion of $\sigma_{\rm \ub}\sim 
10\ \mathrm{km\ s}^{-1}$ 
on the grid resolution scale $\Delta\approx30\ \mathrm{pc}$,
implying turbulent Mach numbers around $10$ in the cold clouds on
the length scale $\ell_{\rm c}$.\footnote{
Note that $\mathcal{M}_{\rm c}$ is different from the Mach number
$\sigma_{\rm \ub}/c_{\rm c}$ scaled down from $\Delta$ to $\ell_{\rm 
c}$ because the latter encompasses both the warm and cold phase.
}
These values are comparable to observational findings 
\citep[e.g.][]{Leroy2008,Shetty12,Stilp2013}. The intensity of small-scale 
turbulence in actively star-forming regions is mainly determined by stellar 
feedback, the exchange of mass between warm and cold phases, and the assumed 
effective pressure balance. Figure~\ref{fig:BS14a_synopsis_K_vs_EffC} shows 
that the models in the 
PN- and HC-families as well as the KM-FK-mff model predict roughly comparable 
star formation efficiencies 
$\epsilon_{\rm mod}\sim 0.1$ (corresponding to star formation time scales of 
$\lesssim100\ \mathrm{Myr}$) 
for the relevant SGS energies around $100\ \kmss$, although their dependence on 
$K_{\rm SGS}$ is quite different. The highest efficiencies are obtained for the 
HC-FK-mff
model, while PN tends to produce relatively low efficiencies. 
This is also reflected by the $\dot{\Sigma}_{\rm SF}$-$\Sigma_{\rm 
H_2}$-relations discussed in section~\ref{sec:sf_relations}. 
Moreover, one can clearly see that the efficiencies following from the KM, 
KM-FK, and PHN models are far below the plausible range of values in turbulent 
star-forming regions. For these models, star formation exclusively occurs in 
regions of exceptionally high density, where efficiencies that are 
systematically too low are compensated by much shorter free-fall time scales.

The star formation efficiencies obtained with the different models can be 
understood as follows:

\subsubsection{PN family}
All PN-family models produce a similar $\epsilon_{\rm mod}$-$K_{\rm SGS}$ 
relation. However, the efficiencies are systematically
lower for the original PN model and the implied star formation time scale
is about $0.75\ \mathrm{Gyr}$, much longer than
$\sim100\ \mathrm{Myr}$ for PN-FK and PN-FK-mff.
Due to the lower level of $\epsilon_{\rm mod}$, the 
disk tends to be clumpier. 
This is mainly a consequence of the chosen model coefficient $\theta=0.5$ 
compared to $\theta=1$ for the FK-calibrated variants of the PN-family (see
Table~\ref{tab:SFR_mods}). 
This behaviour supports our assumption that $\ell_{\rm c}$ should be close
to the integral scale of turbulence in cold gas clouds, but 
the strong sensitivity on the parameter $\theta$ makes
PN-type models difficult to control.

The declining trend of $\epsilon_{\rm mod}$ with increasing SGS turbulence energy 
is due to $\mathrm{exp}(x_{\rm cr})\propto\mathcal{M}^2\propto K_{\rm SGS}$
(see equation~\ref{eq:x_cr_PN}), i.e.\ stronger turbulence tends to produce thinner shock layers, 
requiring higher densities to reach the critical mass. This effect is alleviated,
however, by the implicit dependence of $\rho_{\rm c,pa}$ (the mean density
of the cold phase) on $\mathcal{M}$. This dependence is a consequence of the
effective pressure equilibrium between the phases. Since mostly logarithmic density fluctuations
$x$ close to $x_{\rm cr}$ contribute to the integral over the exponential
tail of $\mathrm{pdf}(x)$ for $x>x_{\rm cr}$, the weighting factor $f_{\rm ff}$ (i.e.\ single vs.\
multi free-fall) has only a minor impact.

\subsubsection{HC family}
All HC-family models are actually multi free-fall models and yield realistic 
$\epsilon_{\rm mod}\sim0.1$. As a 
consequence, the resulting star formation rate is comparable to the PN-FK case, although the 
trend of $\epsilon_{\rm mod}$ with $K_{\rm SGS}$ is opposite to the PN family
because $\mathrm{exp}(x_{\rm cr})\propto\mathcal{M}^{-2}$ (cf. equation~\ref{eq:x_cr_HC} and \ref{eq:x_cr_PN}).
Omitting the second term in the expression for $x_{\rm cr}$ in the case of 
HC-FK-mff (see Table~\ref{tab:SFR_mods}) significantly affects $\epsilon_{\rm mod}$ only 
for values of $K_{\rm SGS}$ that are generally lower than in star-forming regions.

Although the values chosen for the cutoff mass $y_{\rm cut}$ differ
by a factor $10$ in the HC and HC-FK runs, we do not see a significant impact on $\epsilon_{\rm mod}$.
However, this does not necessarily imply that $y_{\rm cut}$ is unimportant.
Probably the influence of $y_{\rm cut}$ is compensated by the calibration
factor $f_{\rm cal}\approx0.2$ in the case of HC-FK compared to $f_{\rm 
cal}=1$ for HC. It is therefore likely that a careful calibration is also
required for HC-type models.

\subsubsection{KM family}
KM-FK-mff differs from PN-FK-mff only by a constant coefficient, which explains the striking similarity of the top right and the bottom right plot in Figure~\ref{fig:BS14a_synopsis_K_vs_EffC}. 
The functional dependence on the Mach number, $\mathrm{exp}(x_{\rm cr})\propto\mathcal{M}^2$, 
is the same for all KM and PN models. However, both models, KM and KM-FK lack a free-fall weighting factor in the integrand: $f_{\rm ff}=1$, independent of $x_{\rm cr}$ (see second column in Table~\ref{tab:SFR_mods}). 
For this reason, $\epsilon_{\rm mod}$ rapidly drops with increasing $K_{\rm 
SGS}\propto\mathcal{M}^2$, which strongly suppresses star formation as soon as
stellar feedback kicks in (about $4\ \mathrm{Myr}$ after creation of the first stars).
The suppressed star formation in turn prevents feedback from becoming sufficiently
strong to expand and destroy a region with active/recent star formation. 
This results gradually growing global star formation rates in the KM and 
KM-FK simulations, without reaching statistically stable disk configurations. 

\subsubsection{PHN-model}
In contrast to the analytical models, \citetalias{Padoan2012} assume a strong increase of 
$\epsilon_{\rm mod}$ with the cold-phase
density through the exponential dependence on the viral parameter. 
This avoids a continuously drifting star formation rate such as in the KM 
simulation. 
Nonetheless, the level of $\epsilon_{\rm mod}$ following from the PHN model is 
systematically too low in the
regime that is relevant for star formation. As we will show in Section~\ref{sc:disk_structure},
this results in an extremely clumpy disk, which only slowly reaches a stable configuration. 
Although the fit formula for $\epsilon_{\rm mod}$ put forward by PHN\footnote{
 This formula is not based on any physical ideas. It is merely a fit to the
 data points obtained from their simulation suite.
}
depends only on the virial parameter $\alpha_{\rm vir}\propto K_{\rm SGS}$, 
the lower envelope of the $\epsilon_{\rm mod}$-$K_{\rm SGS}$-distribution
shown in Fig.~\ref{fig:BS14a_synopsis_K_vs_EffC} suggests
a quadratic dependence on $K_{\rm SGS}$. This non-linear behaviour 
is a consequence of the implicit $K_{\rm SGS}$-dependencies of the cold-gas density  $\rho_{\rm 
c,pa}$ and the clump scale $\ell_{\rm c}$.

\subsection{Disk structure and probability density functions}
\label{sc:disk_structure}

\begin{figure}
\centering
  \includegraphics[width=0.95\hsize]{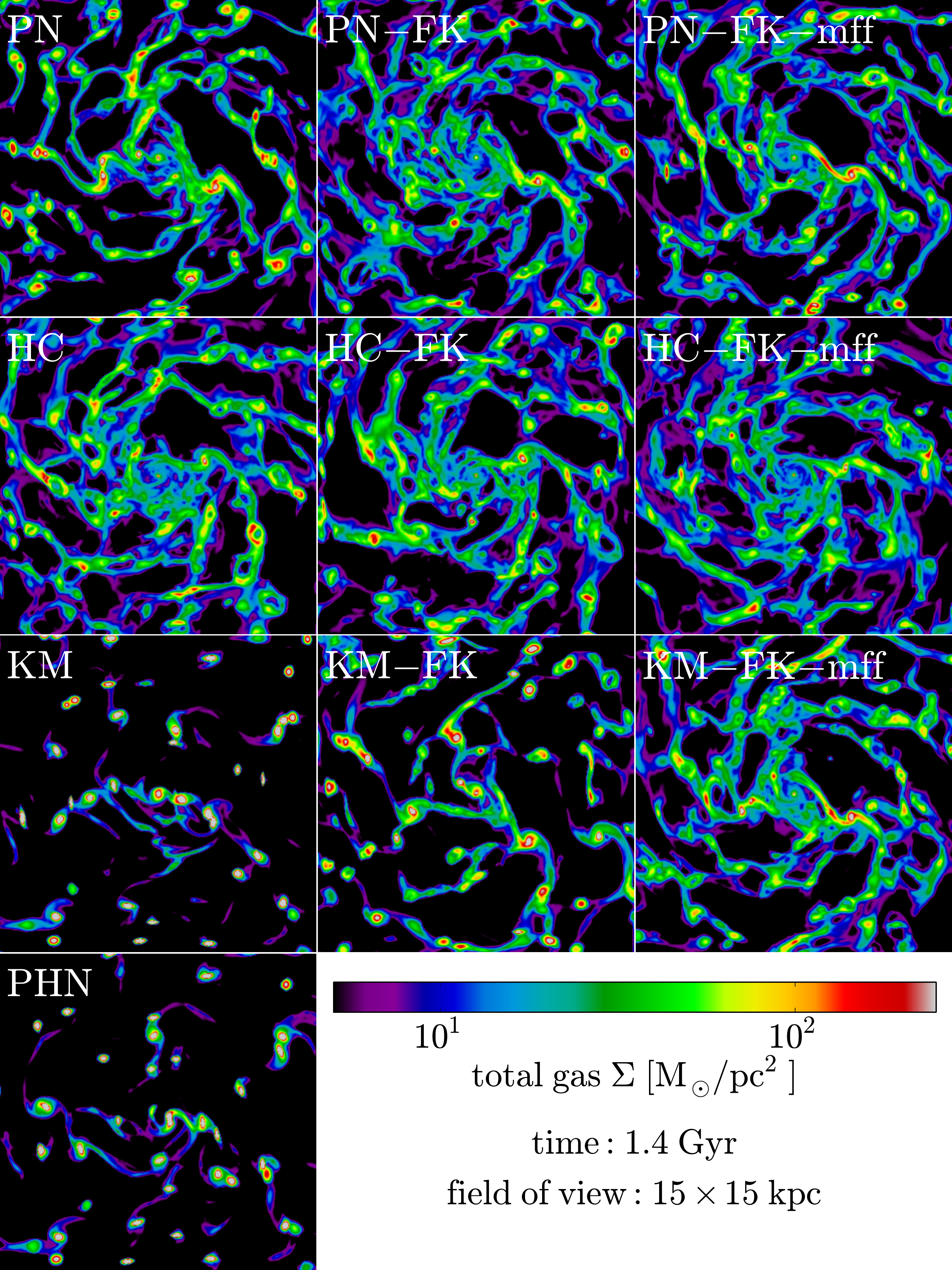}
\caption{Projections of the total gas density $\rho$ perpendicular to the disk 
plane in an area of $15\times15~kpc$ around the galactic centre. The individual 
projections depict the state of the disk in the indicated run at $\sim$1.4~Gyr.}
\label{fig:BS14a_synopsis_SigmaG}
\end{figure}

By comparing the gaseous disk structure for the different runs in 
Fig.~\ref{fig:BS14a_synopsis_SigmaG}, it can be seen that deviations of the 
star formation rate 
caused by different average levels of $\epsilon_{\rm mod}$ 
go along with changes in the disk structure. This is obvious for the cases KM, 
KM-FK, and PHN, which produce much clumpier disks. To a lesser degree, also the 
disk
in the PN run tends to be clumpier as compared to PN-FK. Virtually no 
differences can be seen for the runs from the HC-family, which also compare 
well to PN-FK.
In extreme cases (KM and PHN), the disks are dominated by the presence of very 
massive, and dense clumps 
($\Sigma\gtrsim10^3\ \mathrm{M}_\odot\,\pc^{-2}$).
The void areas ($\Sigma\ll1\ \mathrm{M}_\odot\,\pc^{-2}$) 
between these clumps are only intercepted by tidal tails and bridges that form 
during mergers and fly-by interactions of clusters. As a result, a similar star 
formation rate such as in the late stages of the
PHN and PN-FK runs does not necessarily imply similar disk structures. 

In the KM and PHN runs, we find extremely massive clumps with
very high column densities
($\Sigma\gtrsim10^3\ \mathrm{M}_\odot\,\pc^{-2}$). This is in stark contrast to 
other runs (PN-FK, PN-FK-mff, HC family, and KM-FK-mff), where gas of moderate 
density ($1\ \mathrm{M}_\odot\,\pc^{-2}\lesssim\Sigma\lesssim100\ \mathrm{M}_\odot\,
\pc^{-2}$)
is concentrated in spiral-like structures with substantially smaller clumps,
similar to the initial disk at 1~Gyr.
Denser, star-forming regions are eventually disrupted by the violent expansion
induced by SN feedback, but the low-density gas is less space filling in these runs. 
The disks in the PN and the KM-FK runs are intermediate cases.

\begin{figure}
\centering
  \includegraphics[width=0.95\hsize]{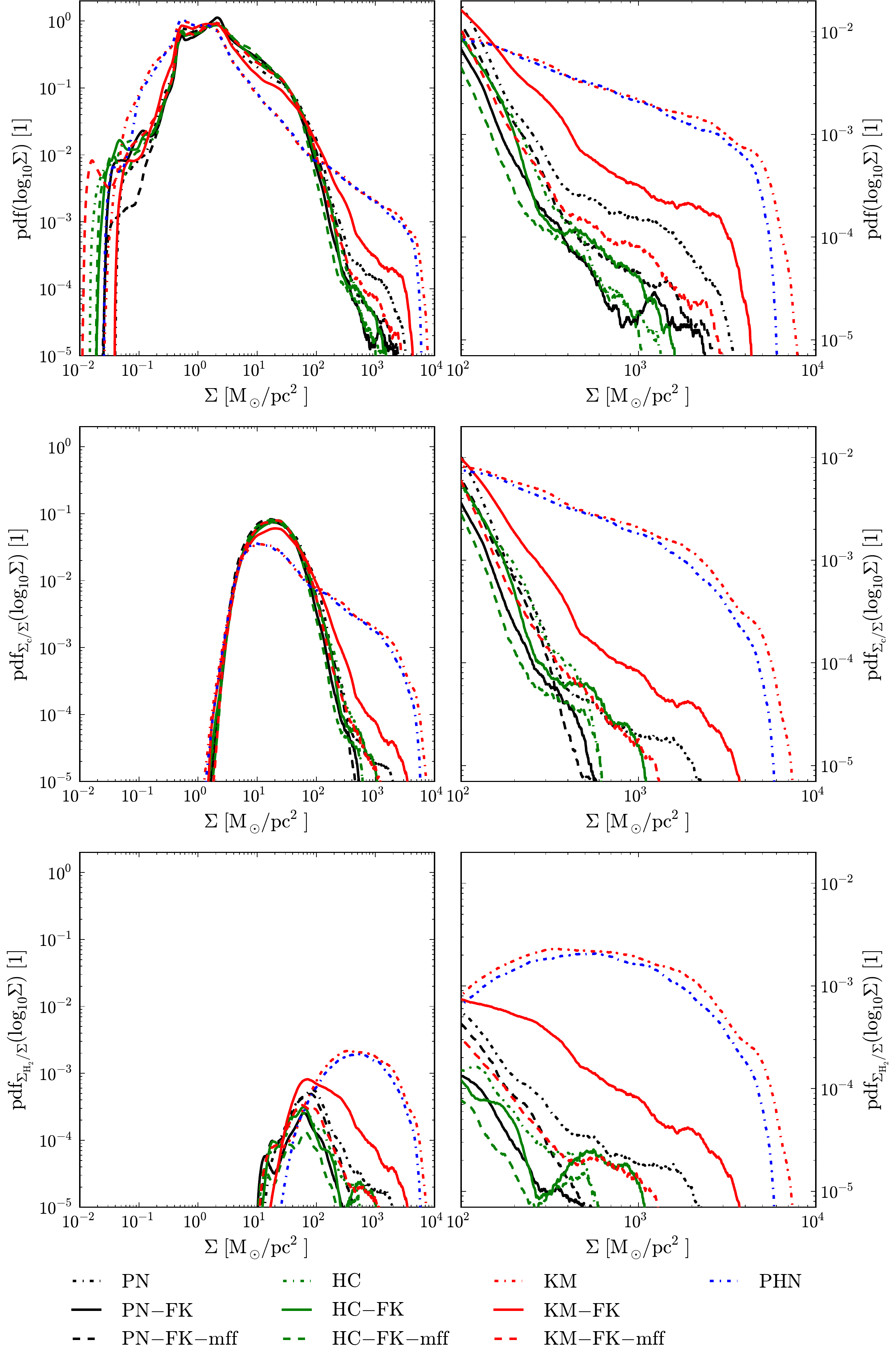}
\caption{\small\textbf{Top left: }Area weighted probability density function 
$\mathrm{pdf}(\log_{10}\Sigma)$ of the logarithmic total gas column density in 
the disks versus total gas column density $\Sigma$ for all runs at a simulated 
age of $\sim1.4$~ Ga. Arrangement of line colours and styles like in 
Fig.~\ref{fig:BS14a_SFRvsTime}. \textbf{Top right: }Detail of the plot on the 
left for $\Sigma>10^2\;\mathrm{M}_\odot \pc^{-3}$. \textbf{Middle left: 
}Probability density function $\mathrm{pdf}_{\Sigma_{\rm 
c}/\Sigma}(\log_{10}\Sigma)$ weighted by the cold phase gas fraction 
$\Sigma_{\rm c}/\Sigma$ versus $\Sigma$. \textbf{Middle right: }Detail of the 
plot on the left for $\Sigma>10^2\;\mathrm{M}_\odot \pc^{-3}$. \textbf{Bottom 
left: }Probability density function $\mathrm{pdf}_{\Sigma_{\rm 
c}/\Sigma}(\log_{10}\Sigma)$ weighted by the shielded molecular gas fraction 
$\Sigma_{\rm H_2}/\Sigma$ versus $\Sigma$.
 \textbf{Bottom right: }Detail of the plot on the left for 
$\Sigma>10^2\;\mathrm{M}_\odot \pc^{-3}$.}
\label{fig:BS14a_pdfs}
\end{figure}

A more accurate distinction is possible by means of probability density 
functions of the gas column density weighted by 
by area or by the mass fraction $\Sigma_{\rm X}/\Sigma$, where $\Sigma_{\rm X}$ 
either stands for the column density of cold gas, $\Sigma_{\rm c}$, or the 
column density of shielded molecular gas, $\Sigma_{\rm H_2}$. The results for 
the different simulation runs are shown in Fig.~\ref{fig:BS14a_pdfs}.
The normalisations of $\mathrm{pdf}[\mathrm{log}_{10}(\Sigma)]$ and 
$\mathrm{pdf}_{\Sigma_{\rm X}/\Sigma}[\mathrm{log}_{10}(\Sigma)]$ are, 
respectively, given by
\begin{equation}
\displaystyle{\int_{-\infty}^{\infty}\mathrm{pdf}[\mathrm{log}_{10}(\Sigma)]\df[
\mathrm{log}_{10}(\Sigma)]=1}\\
\end{equation}
and
\begin{equation}
 \displaystyle{\int_{-\infty}^{\infty}\mathrm{pdf}_{\frac{\Sigma_{\rm 
X}}{\Sigma}}[\mathrm{log}_{10}(\Sigma)]\df[\mathrm{log}_{10}(\Sigma)]=\frac{M_{
\rm X}}{M}}.
\end{equation}
One can distinguish at least four different regimes in the plot of 
$\mathrm{pdf}[\mathrm{log}_{10}(\Sigma)]$ in the upper left panel of 
Fig.~\ref{fig:BS14a_pdfs}:
\begin{enumerate}
 \item $\Sigma\lesssim2\times10^{-1}\ \mathrm{M}_\odot\mathrm{pc}^{-2}$: Gas 
found in this range is extremely hot and dilute and was most likely recently 
produced by SNe.
 \item $2\times10^{-1}\ 
\mathrm{M}_\odot\mathrm{pc}^{-2}\lesssim\Sigma\lesssim3\times10^{0}\ 
\mathrm{M}_\odot\mathrm{pc}^{-2}$: In this range and also at higher densities, 
there is an approximate balance between heating and radiative cooling. The two 
prominent maxima of the pdfs correspond to the collisional ionisation of 
hydrogen at $\approx2\times10^{0}\ \mathrm{M}_\odot\mathrm{pc}^{-2}$ and helium 
at $\approx5\times10^{-1}\ \mathrm{M}_\odot\mathrm{pc}^{-2}$, respectively. For 
this reason, there is hardly any gas in the cold phase below a density of 
$\approx2\times10^{0}\ \mathrm{M}_\odot\mathrm{pc}^{-2}$ (compare top and 
middle plots in Fig.~\ref{fig:BS14a_pdfs}).
 \item $3\times10^{0}\ 
\mathrm{M}_\odot\mathrm{pc}^{-2}\lesssim\Sigma\lesssim10^{2}\ 
\mathrm{M}_\odot\mathrm{pc}^{-2}$: In this regime, the cold and warm phases  
coexist. The distribution of the cold gas shown in the middle plot in 
Fig.~\ref{fig:BS14a_pdfs} has an approximately log-normal shape, with a maximum 
at $\Sigma\approx2\times10^1\ \mathrm{M}_\odot\mathrm{pc}^{-2}$. The existence 
of a cold phase can also be seen as a hump in 
$\mathrm{pdf}[\mathrm{log}_{10}(\Sigma)]$. Above densities of about 
$\approx1\times10^{1}\ \mathrm{M}_\odot\mathrm{pc}^{-2}$, shielded molecular 
gas may be present inside of the cold phase (see bottom plot).
 \item $\Sigma\gtrsim10^{2}\ \mathrm{M}_\odot\mathrm{pc}^{-2}$: Due to 
significant gravitational self-interactions of gas and stars, both 
$\mathrm{pdf}[\mathrm{log}_{10}(\Sigma)]$ and $\mathrm{pdf}_{\Sigma_{\rm 
c}/\Sigma}[\mathrm{log}_{10}(\Sigma)]$ show an excess in their high density 
tails. The gas in this
 regime resides in the prominent, more or less gravitationally bound knots and 
clumps that are visible in Fig.~\ref{fig:BS14a_synopsis_SigmaG}. Shielded 
molecular gas makes up the major fraction of the cold phase, but there is also 
gas in the warm phase because of the evaporation of cold gas due to SN feedback.
\end{enumerate}


Differences induced by the star formation models are most evident in the 
density regime (iv), in which most of the star formation happens. There is an 
enormous excess of high-density gas in case of PHN and KM. The maxima of 
shielded molecular gas are found at roughly ten times higher $\Sigma$ 
($\approx6\times10^{1}\ \mathrm{M}_\odot\mathrm{pc}^{-2}$) compared to the 
maxima in the other runs. The lack of intermediate density gas as well as the 
increased abundance of low-density gas in PHN and KM is readily visible in the 
regimes iii) and iv). The pdfs from the KM-FK and the PN run exhibit 
significantly less pronounced high-density tails, but they are still stronger 
than the tails obtained with the HC-family, PN-FK, PN-FK-mff, and KM-FK-mff 
models.

Observationally almost log-normal shaped pdfs are found for gas in 
nearby galaxies \citep[e.g.][]{Berkhuijsen2015,Hughes2013a}. 
The pdf of CO-line emitting gas in M51 by \citet{Hughes2013a} is based on data of 
comparable spatial resolution (40~pc). Most of what we consider as cold phase gas 
($\Sigma_{\rm c}$) would be seen in such observations, but a significant fraction 
of it is actually atomic, particularly for lower $\Sigma$. Compared to observed
pdfs (e.g. Figure~2a in \citealt{Hughes2013a}), the maximum of 
$\mathrm{pdf}_{\Sigma_{\rm c}/\Sigma}[\mathrm{log}_{10}(\Sigma)]$ 
computed from our simulations is thus shifted toward lower densities. 
Apart from that, M51 is an interacting galaxy in which highly compressed gas 
is expected to be more abundant than in an isolated, quiescent galaxy.
Nevertheless, the general shape and especially the high-density tail as inferred from
the HC-family, PN-FK, PN-FK-mff, and KM-FK-mff model runs are consistent with 
the observations of \citet{Hughes2013a}, while PHN and KM are clearly ruled out.

A comparison with the simulations of \citet{Hopkins2012}
shows pdfs that are populated up to much higher densities. 
This appears to be a consequence of the significantly higher numerical resolution,
but also the star formation recipe used in their simulations. Specifically, the 
low efficiency and the rather high star formation threshold favour the formation of 
denser clouds and a higher global star formation rate (cf.\ their 'Sbc'-runs).

\subsection{Star formation relations}\label{sec:sf_relations}

\begin{figure}
\centering
  \includegraphics[width=0.95\hsize]{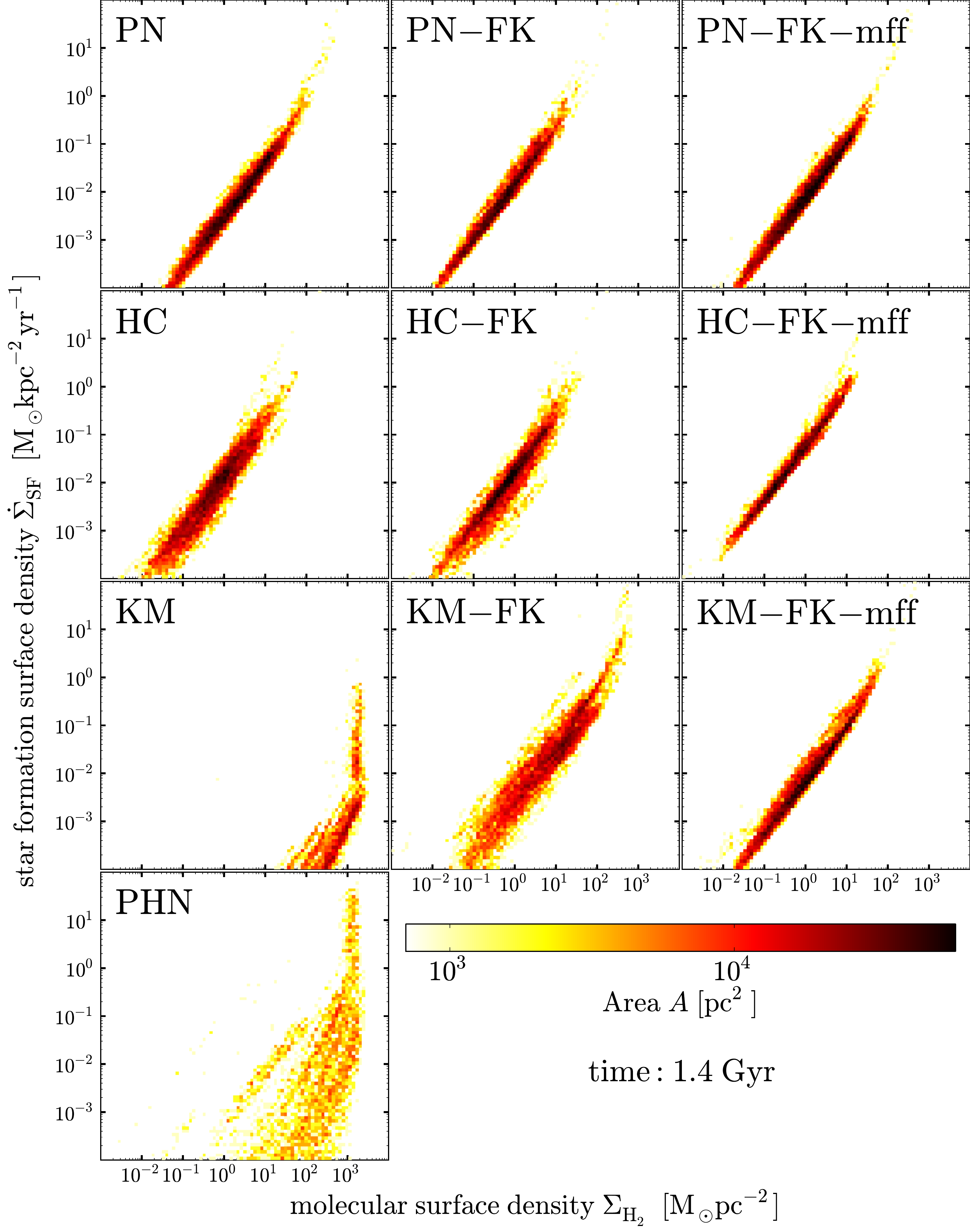}
\caption{Comparison of area-weighted 2D histograms of the star formation column 
density $\dot{\Sigma}_{\rm s}$ vs.\ the column density $\Sigma_{\rm H_2}$ of 
shielded, molecular gas at an age of $\sim1.4$~Gyr. The area is logarithmically 
colour-coded.}
\label{fig:BS14a_synopsis_SigmaSF_vs_SigmaH2}
\end{figure}

With the exception of the KM and PHN runs, a robust relation between the 
star-formation column density $\dot{\Sigma}_{\rm SF}$ and column density of 
shielded, molecular gas $\Sigma_{\rm H_2}$ is established (see 
Fig.~\ref{fig:BS14a_synopsis_SigmaSF_vs_SigmaH2}). However, the inferred time 
scales $\tau_{\rm SF,H_2}$ of star formation in molecular gas vary between 
$\approx80\ \mathrm{Myr}$ and $\approx0.5\ \mathrm{Gyr}$. As pointed out in 
\citetalias{BS14}, 
time scales around $\approx100\ \mathrm{Myr}$ are reasonable because the
simplified model for the molecular fraction $f_{\rm H_2}$ in our simulations is 
biased toward high-density, star-forming regions. 
Observational analogues of what is treated as shielded molecular gas in our simulations
favour short star formation time scales $\sim100\ \mathrm{Myr}$ (e.g.\ 
\citealt{{GaoSol2004,Evans2009,Murray2011}}).
Keeping that in mind, the star formation rate is anomalously low for a wide 
range of $\Sigma_{\rm H_2}$ column densities for the KM- and the PHN-model, 
with a very sharp rise
at $\Sigma_{\rm H_2}\sim 10^3\ \mathrm{M}_\odot\mathrm{pc}^{-2}$. Moreover the 
spread is significantly larger in these cases. Such $\dot{\Sigma}_{\rm 
SF}$-$\Sigma_{\rm H_2}$-relations are definitely not consistent with 
observational counterparts. For the KM-FK and PN models, the relations are 
slightly tilted upwards at the high-density end.\footnote{ 
  This is actually a consequence of a lower bound for the cold-gas fraction at 
densities above $500\ \mathrm{cm}^{-3}$. For the fiducial model, this was 
intended to prevent star formation from being completely quenched by
  feedback in very dense regions, which hardly affected the $\dot{\Sigma}_{\rm 
SF}$-$\Sigma_{\rm H_2}$-relation (see for example, the PN-FK panel in 
Fig.~\ref{fig:BS14a_synopsis_SigmaSF_vs_SigmaH2}). For extremely clumpy disks, 
the star formation rate is artificially pushed upwards. We did not adjust
  the threshold density for the cold-gas floor when carrying out our comparison 
study.}
Basically, the results show that star formation is shifted to higher 
molecular column densities for the models in the KM family, particularly for 
the original KM model, and also for the PHN model. Generally, the multi 
free-fall models produce tighter correlations between $\dot{\Sigma}_{\rm SF}$ 
and $\Sigma_{\rm H_2}$. The tightest correlation is found for HC-FK-mff, with a 
particularly short star formation time scale.

\begin{figure}
\centering
  \includegraphics[width=0.95\hsize]{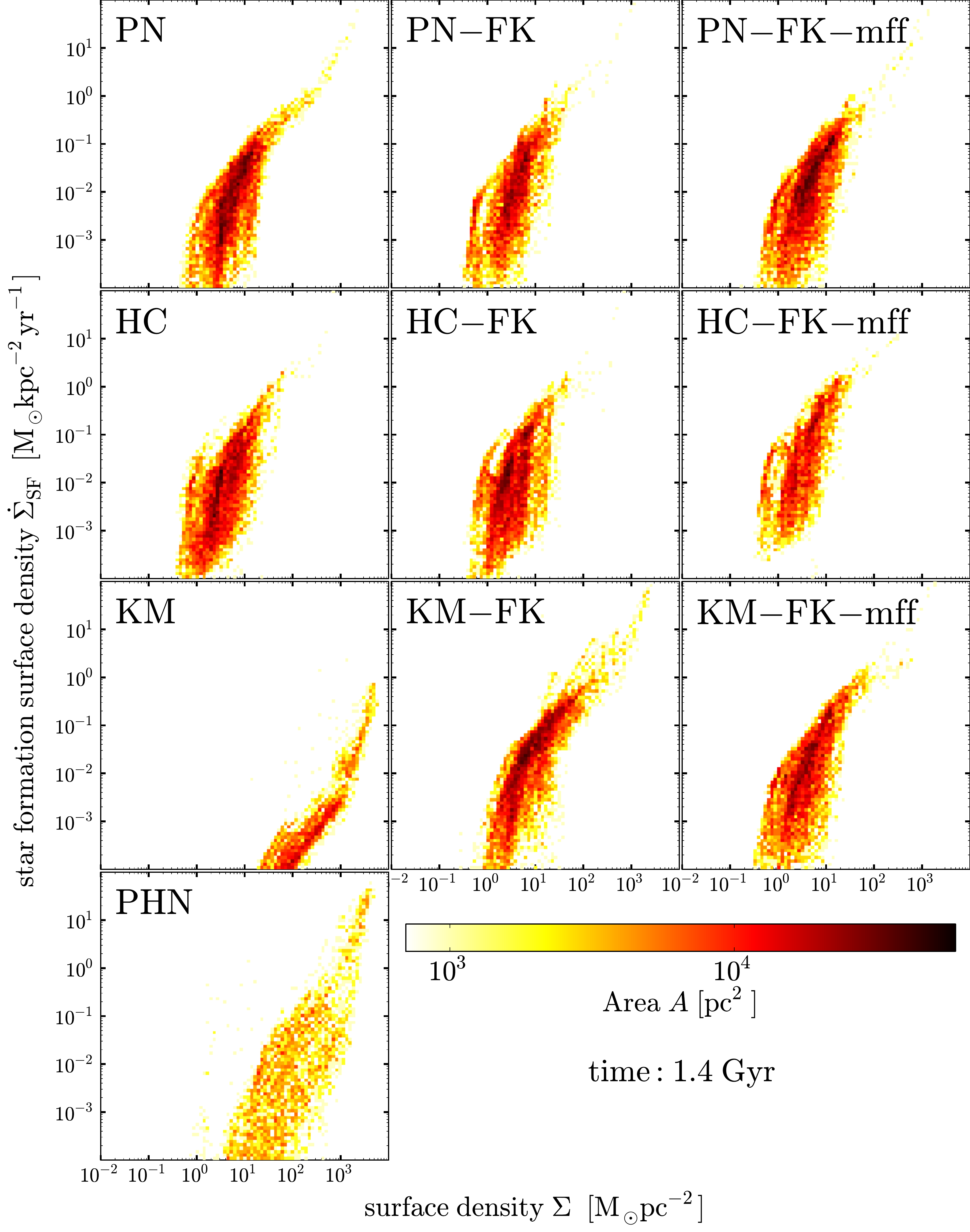}
\caption{2D histograms as in Fig.~\ref{fig:BS14a_synopsis_SigmaSF_vs_SigmaH2} 
for the star formation column density $\dot{\Sigma}_{\rm s}$ vs.\ the total gas 
column density $\Sigma$.}
\label{fig:BS14a_synopsis_SigmaSF_vs_Sigma}
\end{figure}

The 2D histograms of $\dot{\Sigma}_{\rm SF}$ vs.\ $\Sigma$ (total gas column 
density) in Fig.~\ref{fig:BS14a_synopsis_SigmaSF_vs_Sigma} show qualitatively 
the same behaviour as described in \citetalias{BS14}. Star formation is cut off 
below $\Sigma\approx10^0\ \mathrm{M}_\odot\mathrm{pc}^{-2}$ because of the lack 
of shielded, molecular gas. Above this threshold, $\dot{\Sigma}_{\rm SF}$ 
steeply rises and then gradually flattens to an approximate KS-relation 
$\dot{\Sigma}_{\rm SF}\propto\Sigma^\alpha_{\rm KS}$ with $\alpha_{\rm 
KS}\approx2$. An exponent $\alpha_{\rm KS}\approx2$ is indicative of star 
bursts. This is plausible, since the simulated galaxies are still gas-dominated 
after 1.4~Gyr. Again, $\dot{\Sigma}_{\rm SF}$-$\Sigma$-relations obtained with the models KM, PHN, 
and (to a lesser degree) KM-FK differ for the same reasons like in the context of the 
$\dot{\Sigma}_{\rm SF}$-$\Sigma_{\rm H_2}$-relations. 

\section{Discussion and Conclusions}\label{sec:conclude}
In this paper we presented results of a suite of ten isolated disk galaxy 
simulations carried out using the cosmological hydrodynamics code \texttt{Nyx} 
\citep{NYX}, featuring a turbulence SGS model \citep{SchmFed10,Schmidt2014} and 
the MIST model  
\citepalias{BS12,BS14} for the turbulent multi-phase ISM. In each of the runs, 
a different analytic star-formation
model from the literature was applied 
(\citetalias{FedKless12,HC08,Krumholz2005,PadNord09,Padoan2012}, see 
Table~\ref{tab:SFR_mods} for the analytical expressions employed here). These 
models for the sub-resolution structure of the ISM 
were incorporated into the MIST framework to estimate the local star formation efficiency 
$\epsilon_{\rm mod}$ in the cold molecular gas. We used an evolved (for 1~Gyr) disk 
from the 'ref' run of \citetalias{BS14}, 
which resembles a gas-rich, quiescent star-forming spiral galaxy, 
as initial conditions to start the runs with different star formation models. 
This approach minimised effects caused by the rapid growth of stellar mass and 
metallicity during the initial transient. 
From these initial conditions every simulation was evolved over additional 0.4~Gyr, which allows the star 
forming disk to settle into a new equilibrium in most cases.
Several important properties resulting from the local self-regulation in MIST, 
as reported by \citetalias{BS14}, are recovered in our simulation runs 
independent of the choice of the star formation efficiency $\epsilon_{\rm mod}$ 
if the KM, KM-FK, and PHN models are excluded: 
\begin{itemize}
\item A global star formation rate around $\dot{M}_{\rm SF}\approx2.5\ \Msyr$ 
is maintained. This is a typical value for a quiescent isolated
spiral galaxy.
\item Model-dependent variations of $\epsilon_{\rm mod}$ do not significantly 
affect the amount of shielded molecular gas.
	As a result, the almost linear relationship between local star 
formation rate column density $\dot{\Sigma}_{\rm s}$ and column
	density $\Sigma_{\rm H_2}$ of shielded molecular gas turns out to be 
very robust in the framework of MIST. However,  
	the inferred depletion time scales vary between 80~Myr and 0.5~Gyr, 
depending on the star formation model.
	With respect to observational findings, a depletion time scale $<100\ 
\mathrm{Myr}$ is more plausible
	\citep{GaoSol2004,Murray2011}. Moreover, a tighter correlation is 
obtained for the multi free-fall models.
\item Our star-forming regions have a typical life-cycle of about $10$ to $30\ 
\mathrm{Myr}$. A fraction of $\sim10$ per cent of their gaseous mass 
	is converted into stars during that period, which matches observational estimates 
\citep[e.g.][]{Blitz2007,McKee2007,Miura2012} 
	as well as simulation results by e.g.\ \citet{Hopkins2012} and \citet{Tasker2015}. 
We also find long-lived giant star-forming clumps similar to those described by 
\citet{Bournaud2014} in our simulations.
\item Cold-gas density pdfs are comparable to 
	observational counterparts \citep[see, for example,][]{Hughes2013a}, 
particularly at the high densities that are required for star formation.
\item The high-density tails of our $\dot{\Sigma}_{\rm 
s}$-$\Sigma$-distributions suggest a KS-relationship 
	$\dot{\Sigma}_{\rm s}\propto\Sigma^{\alpha_\Sigma}$ with 
$\alpha_\Sigma\approx2$, which is expected for environments dominated by gas
	dynamics.
\item Star formation mainly occurs in the SGS turbulence energy range $10\ 
\kmss\lesssim K_{\rm SGS}\lesssim10^3\kmss$, with a peak 
	around $10^2\kmss$. This value corresponds to a 3D-velocity dispersion 
in the range between $\sigma_{\ub}\approx 3\ \kms$ and $30\ \kms$, 
	which is in good agreement with observations of star-forming regions 
\citep[e.g.][]{Leroy2008,Shetty12,Stilp2013}
	(since these regions are more or less unresolved in our simulations, 
the SGS turbulence energy $K_{\rm SGS}$ is the relevant quantity). 
	A minimum level of turbulence is necessary to boost the molecular 
hydrogen formation rates. On the other side,
	there is an upper bound on $K_{\rm SGS}$ because high values of $K_{\rm 
SGS}$, which are produced by intense stellar feedback, tend to disrupt star 
forming regions. This effect also limits the 
duty cycle of star-forming regions to around 
	$10$ to $30\ \mathrm{Myr}$ in our simulation runs, in agreement with 
observational estimates \citep[e.g.][]{Blitz2007,McKee2007,
	Miura2012}.
\end{itemize}
For most models, $\epsilon_{\rm mod}$ is effectively a function of $K_{\rm 
SGS}$. The scatter of $\epsilon_{\rm mod}$ in 
Fig.~\ref{fig:BS14a_synopsis_K_vs_EffC} indicates a subdominant dependency on 
other quantities such as the density, for example, in the case of PHN. 
For $\epsilon_{\rm mod}\sim 0.1$, corresponding to molecular gas depletion time 
scales of $\lesssim100\ \mathrm{Myr}$ in the simulations, the gaseous star 
forming disks are 
moderately clumpy and exhibit pronounced transient spiral-like, flocculent 
features connecting knots and clumps. 
If $\epsilon_{\rm mod}$ is substantially lower in the interval of turbulence 
energy values that admit significant star formation, the disk structure 
changes. This particularly happens with the KM, KM-FK, and PHN models, which 
produce either too many or not enough stars to support the initial disk 
configuration through feedback. 
In these runs, the disk undergoes a transition to 
a different configuration with markedly more massive and compact clumps
accompanied by tidal tails and bridges. 
Such extremely massive clumps appear to be common in star-forming, 
gas-rich galaxies at high redshifts $\sim2$ 
\citep[e.g.][]{Zanella2015,Guo2012}. Similar structures are found in some
simulations of gas-rich galaxies (e.g. by \citealt{Hopkins2011,Bournaud2014}). 
However, the models considered here are intended to describe conditions in
fairly regular star-forming regions, which are found in almost bulge-less and 
gas-rich spiral galaxies without any external disturber. Some of the models
clearly fail to produce a disk structure that even remotely resembles such galaxies.
However, the strong negative coupling between star formation efficiency and
turbulence produced by stellar feedback, for example, in the case of the KM
model, might help to gain a better understanding of galaxies dominated 
by massive clumps. Which factors actually select
the mode of star formation is left as an open question.

If additional feedback processes 
like winds from young, massive stars were incorporated in our simulations, the 
global star formation rate would be shifted, but the self-regulation mechanism would 
basically remain intact. Apart from that, the star formation rate would be 
altered if the circumgalactic medium were properly taken into account (such as 
in cosmological zoom-in simulations). This has to be investigated in future 
studies. 

Both KM and KM-FK do not feature a free-fall time factor (i.e. $f_{\rm ff}=1$, meaning
that the effective free-fall time scale is given by the mean density of the gas
in cold phase), while all other theoretical models use some weighting factor
to modify the effective free-fall time scale. 
In our simulations, this turns out to be an essential feature to reproduce observed properties  
of nearby disk galaxies and their star-forming regions. 
The importance of a non-trivial free-fall factor, such as in the PN and
the HC models, was also highlighted by \citetalias{FedKless12}.
Moreover, we can confirm that the calibrations established by this study generally
result in improvements.
Although \citetalias{FedKless12} found higher
residual $\chi^2$-values for the best-fit HC models than for the competing models, 
our simulations suggest that the models from the HC family yield the most plausible
results in terms of the correlation between star formation rate and molecular hydrogen
column density, the cold-gas pdf, and the disk structure. However, the differences compared
to the PN-FK and PN-FK-mff runs are minor. Also the KM model produces
consistent results if it is applied in the multi free-fall formulation with the new
calibration of \citetalias{FedKless12}.

In contrast to all other models in our suite, the star formation efficiency in 
the simple PHN model has no
functional dependence on the Mach number and, consequently, is ignorant of the 
fundamentally
different nature of subsonic and supersonic turbulent fluctuations. This does 
not appear plausible, because the 
width of the density pdf is basically determined by the turbulent Mach-number 
in the cold phase $\mathcal{M}$ (see equation~\ref{eq:sigma_x}). 
However, \citetalias{Padoan2012} incorporate magnetic fields, which might also 
influence the star formation
efficiency, although probably to a lesser degree. Possibly, the model could be 
improved by establishing
a fit function from turbulence simulations covering a larger parameter space, 
e.g.\ by including non-solenoidal forcing. 

A further distinction between the models that are so far consistent with the 
observational constraints -- namely, PN-FK, PN-FK-mff, HC, HC-FK, HC-FK-mff, 
and KM-FK-mff -- might be possible on the basis of observational measurements 
of the turbulent velocity
dispersion on molecular-cloud scales and associated star formation 
efficiencies. On the theoretical side, constraints on the relation between star 
formation efficiency and parameters related to turbulence can be obtained from 
high-resolution simulations
of star-forming clouds such as in \citetalias{FedKless12}. 
However, any such simulations make idealised assumptions about
initial and boundary conditions for molecular-cloud turbulence. To check the 
consistency of models for the dynamical
star formation efficiency, it will be necessary to re-simulate a star-forming 
region in global disk galaxy simulation such that
star formation processes are fully resolved and then to compare the results 
with the assumptions that went into the model. 

\section*{Acknowledgements}
We thank Peter Nugent and his team for supporting the development of 
\texttt{Nyx} at the Computational Cosmology Center at LBNL.
In particular, we thank Ann Almgren for her help with the implementation of the 
code components used in this work. 
We also thank Jens Niemeyer for discussions and comments and 
we are grateful for valuable comments by the anonymous referee 
that helped us to improve this paper.
H. Braun was financially supported by the CRC 963 of the German Research 
Council.
The simulations presented in this article were performed on SuperMUC of the LRZ 
(project pr47bi) in Germany. 
We also acknowledge the yt toolkit by \citet{yt} that was used for our analysis 
of numerical data.
\bibliography{paper}
\appendix
\bsp
\label{lastpage}
\end{document}